\documentclass[showpacs,aps,twocolumn]{revtex4}
\usepackage{epsf}
\newcommand{\mb}[1]{ { \mbox{\boldmath{$#1$}}}  } 
 
\begin{document}
    
\title{Decoherence effect on the Fano lineshapes in double quantum dots\\ 
        coupled between normal and superconducting leads}


\author{J.\ Bara\'nski  and  T.\ Doma\'nski}
\affiliation{
       Institute of Physics, M.\ Curie-Sk\l odowska University, 
       20-031 Lublin, Poland}

\date{\today}

\begin{abstract}
We investigate the Fano-type spectroscopic lineshapes of the T-shape 
double quantum dot coupled between the conducting and superconducting 
electrodes and analyze their stability on a decoherence. Because of
the proximity effect the quantum interference patterns appear 
simultaneously at $\pm \varepsilon_{2}$, where $\varepsilon_{2}$ 
is an energy of the side-attached quantum dot. We find that decoherence 
gradually suppresses both such interferometric structures. We also 
show that at low temperatures another tiny Fano-type structure can be 
induced upon forming the Kondo state on the side-coupled quantum dot 
due to its coupling to the floating lead.

\end{abstract}

\maketitle

\section{Introduction}
When nanoscopic objects such as the quantum dots, nonowires or thin 
metallic layers are placed in a neighborhood of superconducting material 
they partly absorb its order parameter. On a microscopic level this 
proximity effect causes that electrons near the Fermi energy become bound 
into pairs. Upon forming a circuit with external leads (which can be 
chosen as conducting, ferromagnetic or superconducting) such effect 
can induce a number of unique properties in the normal and anomalous 
tunneling channels \cite{Rodero-11}. For instance, the relation 
between correlations and the on-dot induced pairing has been 
recently experimentally probed by the Andreev spectroscopy 
\cite{Deacon_etal,Pillet-10} and the Josephson current measurements 
\cite{Maurand-12,Novotny-06,Dam-06,Cleuziou-06} signifying important 
role of the Kondo effect on the subgap current. 

We address here the Andreev-type transport through the double 
quantum dot (DQD) nanostructure coupled between the normal (N) 
and superconducting (S) leads. We focus on the subgap regime, 
i.e.\ energies considerably smaller than the pairing gap 
$\Delta$ of superconductor.  Under such conditions  eigenstates 
of the uncorrelated quantum dots are represented either by the singly 
occupied states $\left| \uparrow \right>$, $\left| \downarrow \right>$ 
or by coherent superpositions of the empty and doubly occupied 
configurations $u \left| 0 \right> + v \left|\uparrow\downarrow 
\right>$. The resulting Bogolubov-type quasiparticle excitations 
have an influence on additional spectroscopic features originating 
for instance from the internal structure, the correlations,  
perturbations etc. Due to the proximity effect all these appearing 
structures would show up {\em simultaneously} at negative 
and at positive energies. 

\begin{figure}
\epsfxsize=9cm\centerline{\epsffile{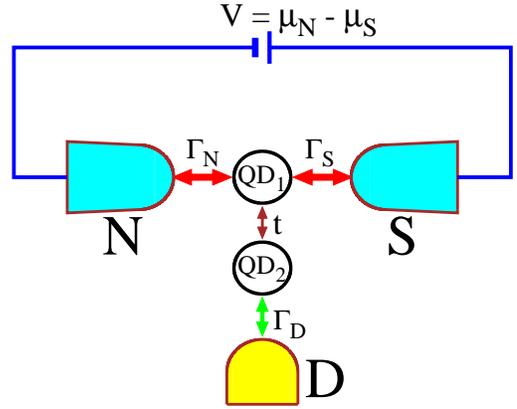}}
\vspace{-0.7cm}
\caption{(color online) Schematic view of the T-shape double quantum 
dot coupled to normal (N) and superconducting (S) electrodes  and in
addition affected by the floating (D) lead that provides a decoherence.}
\label{scheme}
\end{figure}

To highlight this sort of emerging physics we shall explore 
in more detail the interference patterns originating from a charge 
leakage $t$ (assumed to be much weaker than $\Gamma_{N}$ and $\Gamma_{S}$) 
between the central quantum dot (QD$_{1}$) and another side-attached one 
(QD$_{2}$). We also analyze stability of these patterns with respect 
to a decoherence induced by coupling $\Gamma_{D}$ to the floating lead (D) 
as sketched in figure \ref{scheme}. Practically this D electrode can 
mean a substrate on which the quantum dots are deposited or it
mimics the effects caused by phonons/photons \cite{Gao-08}.

Without a decoherence the T-shape double quantum dot systems have 
been already studied theoretically considering both metallic leads 
(see e.g.\ \cite{Zitko-10}) and metallic/superconducting ones \cite{Tanaka-08,
Yamada-10,Baranski-11}. In the regime of weak interdot coupling $t$ this 
configuration of the quantum dots enables realization of the Fano-type 
lineshapes (for a survey on the Fano effect and its realizations in 
various systems see Ref.\ \cite{Miroshnichenko-10}). These features 
can arise when the electron waves transmitted between the external 
electrodes via a broad QD$_{1}$ spectrum happen to interfere with 
the other electron waves resonantly scattered by the discrete QD$_{2}$ 
levels \cite{Trocha-12}. The hallmarks of destructive/constructive 
quantum interference show up in a form of the asymmetric lineshapes 
$G_{0}\frac{\left( x + q \right)^{2}}{x^{2}+1}+G_{1}$ in the tunneling 
conductance, where the dimensionless argument $x$ is proportional 
to $eV - \varepsilon_{2}$, $q$ denotes the asymmetry parameter and 
$G_{0,1}$ are some background functions slowly varying with respect 
to $V$. Such lineshapes have been indeed observed experimentally 
for the DQD coupled between the metallic leads \cite{Sasaki-09,
Kobayashi-04}. Similar Fano-type features have been also previously 
reported from the spectroscopic measurements for a number of systems, 
e.g.\ the cobalt adatoms deposited on Au(111) surfaces \cite{Madhavan-01}, 
the semiopen nanostructures \cite{Fuhner-02,semiopen_theor}, 
the dithiol benzene molecule placed between the gold electrodes 
\cite{Grigoriev-06}, the 'hidden order' phase of the heavy fermion 
compound URu$_{2}$Si$_{2}$ \cite{Schmidt-10}, the dopant atoms 
located in the metal near a Schottky barrier MOSFET \cite{Calvet-11}, 
and many other \cite{Miroshnichenko-10}.     

Considering the proximity effect in N-DQD-S heterojunctions we have 
recently emphasized \cite{Baranski-11} the possibility to observe the 
particle/hole Fano-type lineshapes in the subgap Andreev transport. 
We would like to explore here how such Fano-type structures are robust 
on a decoherence. Since the floating lead (D) does not belong to a closed 
circuit we shall assume that a net current to/from such electrode vanishes, 
so its role can be treated merely as the source of a decoherence. 
Formally our study extends the previous results of Ref.\ \cite{Gao-08} 
onto the anomalous Andreev transport. To our knowledge such problem 
has not been yet addressed in the literature and it might be of practical 
importance for the possible experimental measurements. Influence of 
the bosonic (phonon/photon) modes shall be discussed elsewhere.

In the next section we briefly state formal aspects of the problem. 
Next, we discuss a changeover of the Fano-type lineshapes with respect to 
the asymmetric coupling $\Gamma_{S}/\Gamma_{N}$ which controls efficiency 
of the proximity effect. We also investigate in detail stability of the
particle/hole Fano features with respect to decoherence (in the spectrum 
and in the Andreev transmittance). Finally, we take into account the correlations. 
In particular we argue that for strong enough coupling $\Gamma_{D}$ the Kondo 
resonance formed on the side-attached quantum dot QD$_{2}$ can induce a tiny 
interferometric pattern at $\omega=0$. Such Kondo driven Fano structure
could be detectable by the low bias Andreev conductance.

\section{Theoretical formulation}

The double quantum dot nanostructure shown in Fig.\ \ref{scheme} 
can be described by the following Anderson impurity Hamiltonian
\begin{eqnarray} 
\hat{H} =   \hat{H}_{bath} + \hat{H}_{DQD} + \hat{H}_{T}
\label{model} 
\end{eqnarray}
where the bath $\hat{H}_{bath}=\sum_{\beta}\hat{H}_{\beta}$ consists of 
three external charge reservoirs  ($\beta\!=\!N,S,D$), $\hat{H}_{DQD}$  
refers to the double quantum dot, and $\hat{H}_{T}$ stands for the   
hybridization part. We treat the conducting leads ($\beta\!=\!N,D$) as  
free Fermi gas $\hat{H}_{\beta}=\sum_{{\bf k},\sigma} \xi_{{\bf k}\beta} 
\hat{c}_{{\bf k} \sigma \beta}^{\dagger} \hat{c}_{{\bf k}\sigma \beta}$ 
and represent the isotropic superconductor by the bilinear BCS form $\hat{H}_{S} \!=\!
\sum_{{\bf k},\sigma}  \xi_{{\bf k}S} \hat{c}_{{\bf k} \sigma S }^{\dagger}  
\hat{c}_{{\bf k} \sigma S} \!-\! \Delta \sum_{\bf k} ( \hat{c}_{{\bf k} 
\uparrow S }^{\dagger} \hat{c}_{-{\bf k} \downarrow S }^{\dagger} \!+\! 
\hat{c}_{-{\bf k} \downarrow S} \hat{c}_{{\bf k} \uparrow S})$. Using the 
second quantization we denote by $\hat{c}_{{\bf k} \sigma \beta} 
^{({\dagger})}$ the annihilation (creation) operators for  spin $\sigma
=\uparrow,\downarrow$ electrons in the momentum state ${\bf k}$ with the 
energy $\xi_{{\bf k}\beta}\!=\!\varepsilon_{{\bf k}\beta} \!-\!\mu_
{\beta}$ measured with respect to the chemical potential $\mu_{\beta}$.

Following Ref.\ \cite{Gao-08} we assume that the charge transport 
occurs through the T-shape configuration (Fig.\ \ref{scheme}) only 
via the central ($i\!=\!1$) quantum dot, whereas the side-attached 
quantum dot is responsible merely for the quantum interference. 
Hybridization of the quantum dots with external reservoirs of the 
charge carriers is given by
\begin{eqnarray} 
\hat{H}_{T} &=& \sum_{\beta = N,S} \sum_{{\bf k},\sigma} 
\left( V_{{\bf k} \beta} \; \hat{d}_{1 \sigma}^{\dagger}  
\hat{c}_{{\bf k} \sigma \beta } + \mbox{\rm H.c.} \right) 
\nonumber \\
& & + \sum_{{\bf k},\sigma} 
\left( V_{{\bf k} D} \; \hat{d}_{2 \sigma}^{\dagger}  
\hat{c}_{{\bf k} \sigma D} + \mbox{\rm H.c.} \right)
\label{hybr}
\end{eqnarray} 
Such couplings indirectly affect the quantum dots 
\begin{eqnarray} 
 \hat{H}_{DQD} &=& \sum_{\sigma,i} \varepsilon_{i}  
\hat{d}^{\dagger}_{i \sigma} \hat{d}_{i \sigma} +
t \sum_{\sigma} \left(  \hat{d}_{1\sigma}^{\dagger}  
\hat{d}_{2\sigma} \!+\! \mbox{\rm H.c.} \right)
\\ & & + \sum_{i}U_{i} \; 
\hat{d}^{\dagger}_{i \uparrow} \hat{d}_{i \uparrow} \; 
\hat{d}^{\dagger}_{i \downarrow} \hat{d}_{i \downarrow}  
\label{DQD} 
\nonumber
\end{eqnarray} 
through the interdot hopping $t$ in $\hat{H}_{DQD}$.
We use standard notation for the annihilation (creation) 
operators $\hat{d}_{i}^{({\dagger})}$ for electrons in both
quantum dots $i\!=\!1,2$. Their energy levels are denoted 
by $\varepsilon_{i}$ and $U_{i}$ stand for the on-dot 
Coulomb potential.

If the chemical potentials $\mu_{\beta}$ in the electrodes are 
safely distant from the band edges one can impose the wide-band 
limit approximation, introducing the constant couplings 
$\Gamma_{\beta}=2\pi\sum |V_{{\bf k} \beta}|^{2}\delta\left( 
\omega\!-\!\xi_{{\bf k}\beta}\right)$. In this work we shall 
use $\Gamma_{N}$ as a convenient unit for the energies.

\section{Particle-hole Fano lineshapes}

In order to account for the proximity effect we have to deal 
with the mixed particle and hole degrees of freedom. Among the 
possible ways for doing this one can use the Nambu spinor notation 
$\hat{\Psi}_{j}^{\dagger}\!=\!(\hat{d}_{j\uparrow}^{\dagger},
\hat{d}_{j\downarrow})$ and $\hat{\Psi}_{j}\!=\!(\hat{\Psi}_{j}
^{\dagger})^{\dagger}$. The spectroscopic and transport properties 
of the setup can be determined from the matrix Green's 
function ${\mb G}_{j}(t,t_{0})\!=\!-i\hat{T}\langle \hat{\Psi}_{j}
(t)\hat{\Psi}_{j}^{\dagger}(t_{0})\rangle$. In equilibrium 
case this function depends solely on the time difference $t-t_{0}$ 
and its Fourier transform can be expressed by the following Dyson 
equation
\begin{eqnarray} 
{\mb G}_{j}(\omega)^{-1} = {\mb g}_{j}(\omega)^{-1}
- {\mb \Sigma}_{j}^{0}(\omega)  
- {\mb \Sigma}_{j}^{e-e}(\omega)  ,
\label{GF}\end{eqnarray} 
where ${\mb g}_{j}(\omega)$ are the Green's functions of the 
isolated quantum dots 
\begin{eqnarray} 
{\mb g}_{j}(\omega) = \left( \begin{array}{cc}  
\frac{1}{\omega-\varepsilon_{j}} & 0 \\ 
0 &  \frac{1}{\omega+\varepsilon_{j}}
\end{array} \right)
\label{g_2}
\end{eqnarray} 
and the selfenergies consist of the noninteracting part
${\mb \Sigma}_{j}^{0}(\omega)$ with the additional 
correction ${\mb  \Sigma}_{j}^{e-e}(\omega)$ due to 
the electron-electron correlations.

\begin{figure}
\epsfxsize=10cm\centerline{\epsffile{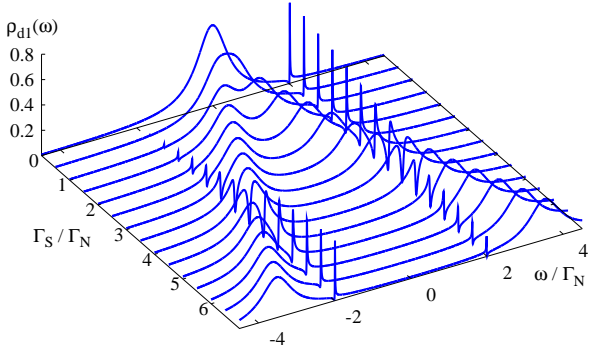}}
\caption{(color online) Particle and hole Fano-type lineshapes 
appearing at $\pm \varepsilon_{2}$ in the spectral function 
$\rho_{d1}(\omega)$ of the central quantum dot. Calculations are 
done for the following parameters $\varepsilon_{1}=0$, 
$\varepsilon_{2}=2\Gamma_{N}$, $U_{i}=0$, $t=0.2\Gamma_{N}$ 
and $\Delta = 10 \Gamma_{N}$.}
\label{dos_evol}
\end{figure}

In the simplest manner a development of the particle and hole 
interference Fano structures (see Fig.\ \ref{dos_evol}) can be
explained restricting to the noncorrelated quantum dots. The 
selfenergies $\mb{\Sigma}_{j}^{0}(\omega)$ are given by 
\begin{eqnarray}
\mb{\Sigma}_{j}^{0}(\omega) = \!\!\! \sum_{{\bf k},\beta} 
V_{{\bf k}\beta} \;\; {\mb g}_{\beta}({\bf k},\omega) \; 
V_{{\bf k}\beta}^{*} + t \; {\mb G}_{j'}(\omega) \; t^{*} ,
\label{S_r}
\end{eqnarray}
where the inderdot hopping contribution refers to $j'\neq j$.
The Green's functions of the conducting leads $\beta=N,D$ 
have diagonal form
\begin{eqnarray} 
{\mb g}_{\beta}({\bf k}, \omega) = 
\left( \begin{array}{cc}  
\frac{1}{\omega-\xi_{{\bf k}\beta}} & 0 \\ 
0 &  
\frac{1}{\omega+\xi_{{\bf k}\beta}}
\end{array}\right)
\label{gN}
\end{eqnarray} 
whereas the superconducting lead is given by the BCS structure 
\begin{eqnarray} 
{\mb g}_{S}({\bf k}, \omega) = 
\left( \begin{array}{cc}  
\frac{u^{2}_{\bf k}}{\omega-E_{\bf k}}+\frac{v^{2}_{\bf k}}
{\omega+E_{\bf k}} \hspace{0.2cm} & \frac{-u_{\bf k}v_{\bf k}}
{\omega-E_{\bf k}}+\frac{u_{\bf k}v_{\bf k}}{\omega+E_{\bf k}}
\\ 
\frac{-u_{\bf k}v_{\bf k}}{\omega-E_{\bf k}}+
\frac{u_{\bf k}v_{\bf k}}{\omega+E_{\bf k}}
& \frac{u^{2}_{\bf k}}{\omega+E_{\bf k}}+
\frac{v^{2}_{\bf k}}{\omega-E_{\bf k}}
\end{array}\right) 
\label{gS}\end{eqnarray} 
with the corresponding coefficients 
\begin{eqnarray} 
u^{2}_{\bf k},v^{2}_{\bf k} &=& \frac{1}{2} \left[ 1 \pm 
\frac{\xi_{{\bf k}S}}{E_{\bf k}} \right]
\nonumber \\
u_{\bf k}v_{\bf k} &=& \frac{\Delta}{2E_{\bf k}} ,
\nonumber
\end{eqnarray}
and the quasiparticle energy $E_{\bf k}\!=\!
\sqrt{\xi_{{\bf k}S}^{2}+\Delta^{2}}$.

In the wide-band limit we obtain for $\beta=N,D$  
\begin{eqnarray}
\sum_{{\bf k}} V_{{\bf k}\beta} \;\; {\mb g}_{\beta}({\bf k},\omega) 
\; V_{{\bf k}\beta}^{*} =  -i \frac{\Gamma_{\beta}}{2} \; 
\left( \begin{array}{cc}  
1 & 0 \\ 0 & 1 \end{array} \right)
\label{Sigma_N}
\end{eqnarray} 
and for the superconducting electrode
\begin{eqnarray}
\sum_{{\bf k}} V_{{\bf k}S} \; {\mb g}_{S}({\bf k},\omega) 
\; V_{{\bf k}S}^{*} = -i \frac{\Gamma_{S}}{2} \gamma(\omega)
\left( \begin{array}{cc}  
1 & \frac{\Delta}{\omega} \\ 
 \frac{\Delta}{\omega}  & 1 
\end{array} \right)
\label{Sigma_S}
\end{eqnarray} 
with
\begin{eqnarray}
\gamma(\omega) = 
\frac{|\omega| \; \Theta(|\omega|\!-\!\Delta)}
{\sqrt{\omega^{2}-\Delta^{2}}}
-\frac{i \omega \; \Theta(\Delta\!-\!|\omega|)}
{\sqrt{\Delta^{2}-\omega^{2}}} .
\label{gamma}
\end{eqnarray} 
In a far subgap regime $|\omega|\ll \Delta$ only the off-diagonal 
terms of the matrix (\ref{Sigma_S}) are preserved tending to 
the static value $-\Gamma_{S}/2$. This {\em atomic limit} case 
has been studied by several groups and the results have been 
recently summarized in the Ref.\ \cite{Yamada-11}. For arbitrary 
$\Delta$ we obtain the following set of coupled equations
\begin{eqnarray}
\mb{G}_{1}(\omega)^{-1} &=& 
\left[ \omega +i \; \frac{\Gamma_{N}+\gamma(\omega)
\Gamma_{S}}{2} \right] {\mb I} 
\!-\!\varepsilon_{1} {\mb \sigma}_{z} \nonumber \\ &&
+ \; i \; \frac{\gamma(\omega) \Delta \Gamma_{S}}{2\omega} 
{\mb \sigma}_{y} - |t|^{2} \; {\mb G}_{2}(\omega) ,
\label{S_QD1}
\\
\mb{G}_{2}(\omega)^{-1} & = & 
\left[ \omega +i \; \frac{\Gamma_{D}}{2} \right] {\mb I} 
\!-\!\varepsilon_{2} {\mb \sigma}_{z} 
-  |t|^{2} \; {\mb G}_{1}(\omega)  
\label{S_QD2} 
\end{eqnarray}
where ${\mb I}$ stands for the identity matrix and 
${\mb \sigma}_{y,z}$ denote the usual Pauli matrices.

\begin{figure}
\epsfxsize=8cm\centerline{\epsffile{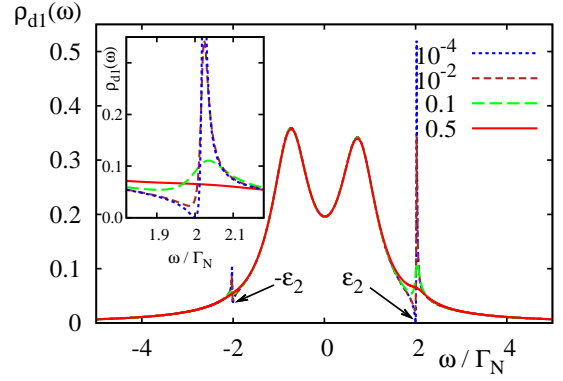}}
\vspace{-0.5cm}
\epsfxsize=8cm\centerline{\epsffile{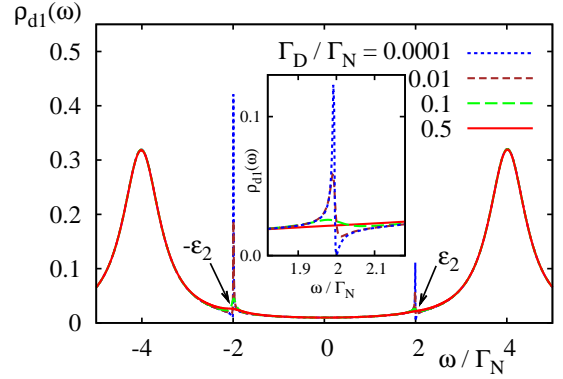}}
\caption{(color online) Spectral function $\rho_{d1}(\omega)$ of the 
central quantum dot in the equilibrium situation. The upper panel 
corresponds to $\Gamma_{S}=1.5\Gamma_{N}$ (when the quasiparticle 
energy $E_{d1}<\varepsilon_{2}$) while the lower one refers to 
$\Gamma_{S}=8\Gamma_{N}$ (when $E_{d1}>\varepsilon_{2}$). We used 
for computations the model parameters  $\varepsilon_{1}=0$, 
$\varepsilon_{2} \!=\!2 \Gamma_{N}$, $t=0.2 \Gamma_{N}$, $U_{i}=0$ 
and several values of $\Gamma_{D}$.}
\label{decoh_dos}
\end{figure}

Figure \ref{dos_evol} shows the spectral function $\rho_{d1}(\omega)$ 
obtained in the equilibrium situation for both uncorrelated quantum 
dots ($U_{i}\!=\!0$) assuming a weak interdot hopping $t=0.2\Gamma_{N}$ 
(decoherence is not taken into account here). 
To focus on the subgap regime $|\omega| \ll \Delta$ we used $\Delta
=10\Gamma_{N}$ and other effects related to the gap edge singularities  
are saparately discussed in the appendix A.
For an increasing ratio $\Gamma_{S}/\Gamma_{N}$ we can notice 
the following qualitative changes: 
a) the initial lorentzian centered at $\varepsilon_{1}$ 
splits into two quasiparticle peaks centered at $\pm E_{1} 
\simeq \pm \sqrt{\varepsilon_{1}+(\Gamma_{S}/2)^{2}}$ 
(due to the proximity effect), 
b) the usual Fano-type lineshape formed at $\varepsilon_{2}$ 
is for larger values of $\Gamma_{S}$ accompanied by appearance
of its mirror reflection at $-\varepsilon_{2}$ (we shall refer 
to these peaks as the particle/hole Fano structures), c) 
Fano-type lineshapes of these particle/hole features are 
characterized by an opposite sign of the asymmetry parameter 
$q$, d)  the asymmetry parameters exchange the sign for such 
$\Gamma_{S}$ when the quasiparticle energy 
$\sqrt{\varepsilon_{1}^{2}+(\Gamma_{S}/2)^{2}} \!\sim\! 
\varepsilon_{2}$.

For a closer inspection on the above mentioned changes we examine 
in the upper (bottom) panel of Fig.\ \ref{decoh_dos} the spectral 
function $\rho_{d1}(\omega)$  obtained for $\Gamma_{S}/\Gamma_{N}=1.5$ 
($8$) when $\varepsilon_{2}$ is smaller (larger) than the quasiparticle 
energy $E_{1}$. We also check the decoherence effect
on these particle and hole Fano lineshapes. We notice that already
a weak coupling $\Gamma_{D}$ to the floating lead washes out both 
these particle and hole Fano structures. Thus we conclude that 
decoherence has a detrimental effect on the quantum interferometric 
features. To provide some physical argumentation for this behavior 
let us recall that the resonant level at $\varepsilon_{2}$ gradually 
broadens uppon increasing $\Gamma_{D}$. For this reason the electron 
waves are scattered on the side-attached quantum dot without any 
sharp change of the phase, thereby the Fano-type interference is 
no longer possible \cite{Trocha-12}. In other words, the 
particle/hole Fano-type lineshapes seem to be rather fragile 
entities with respect to $\Gamma_{D}$. This remark should be 
taken into account by experimentalists while constructing the 
double quantum dot structures on a given substrate material.

\section{Andreev spectroscopy}

Any practical observation of the interferometric particle/hole Fano 
lineshapes could be detectable only in the tunneling spectroscopy. 
For this purpose one could measure e.g.\ the differential conductance 
at small bias (i.e.\ in the subgap regime $|eV|<\Delta$) when charge 
transport is provided solely via the anomalous Andreev current 
$I_{A}(V)$. Skipping the details we apply here the popular
Landauer-type expression  
\begin{eqnarray} 
I_{A}(V) = \frac{2e}{h} \int d\omega T_{A}(\omega)
\left[ f(\omega\!-\!eV,T)\!-\!f(\omega\!+\!eV,T)\right],
\label{I_A}
\end{eqnarray} 
derived previously in the Refs \cite{Krawiec-04,Sun-99}. The 
Andreev current depends on occupancy  $f(\omega\! \pm \!eV,T)$
of the conducting lead (N) convoluted with the transmittance 
$T_{A}(\omega)$. The latter quantity can be determined from
the off-diagonal part of the retarded Green's function 
${\mb G}_{1}(\omega)$ via \cite{Sun-99,Domanski-08}
\begin{eqnarray} 
T_{A}(\omega)=\Gamma_{N}^{2} \left| G_{1,12}(\omega) 
\right|^{2} .
\label{Transmittance_A}
\end{eqnarray} 
The Andreev transmittance (\ref{Transmittance_A}) is a dimensionless 
quantity and, roughly speaking, it is a measure of the proximity 
induced on-dot pairing. Of course (\ref{Transmittance_A}) depends 
indirectly on various structures appearing in the spectrum of the 
central quantum dot, including the particle-hole Fano features. 

\begin{figure}
\epsfxsize=8cm\centerline{\epsffile{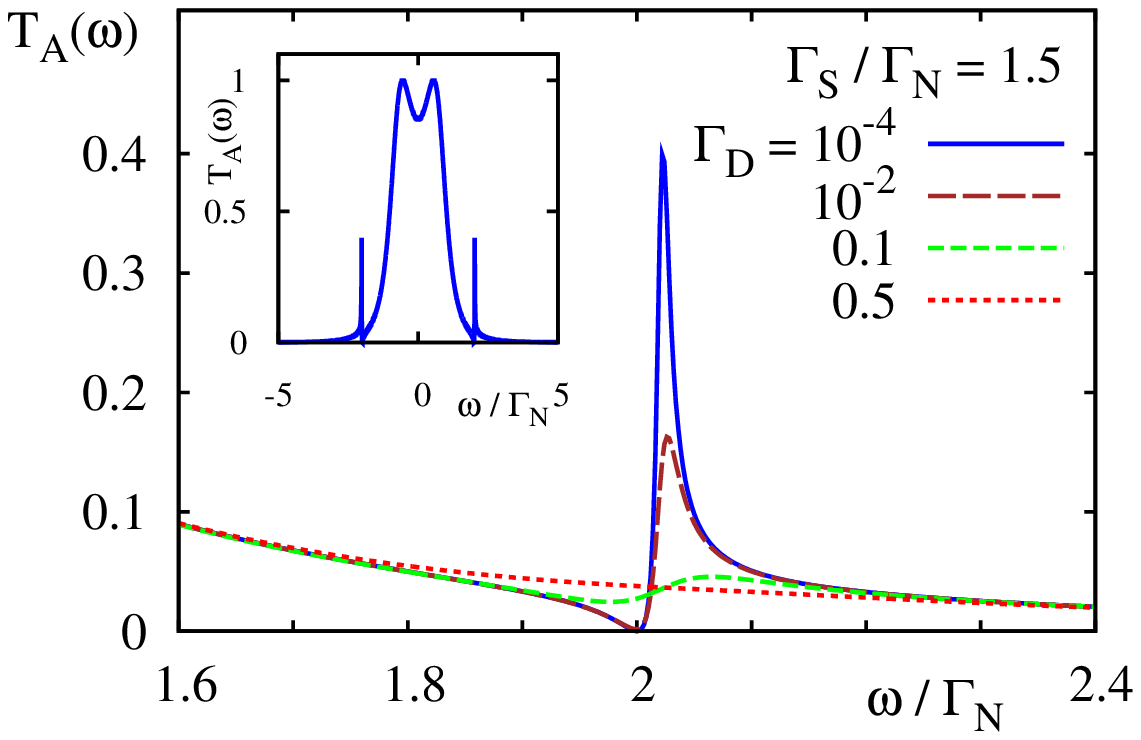}}
\vspace{-0.5cm}
\epsfxsize=8cm\centerline{\epsffile{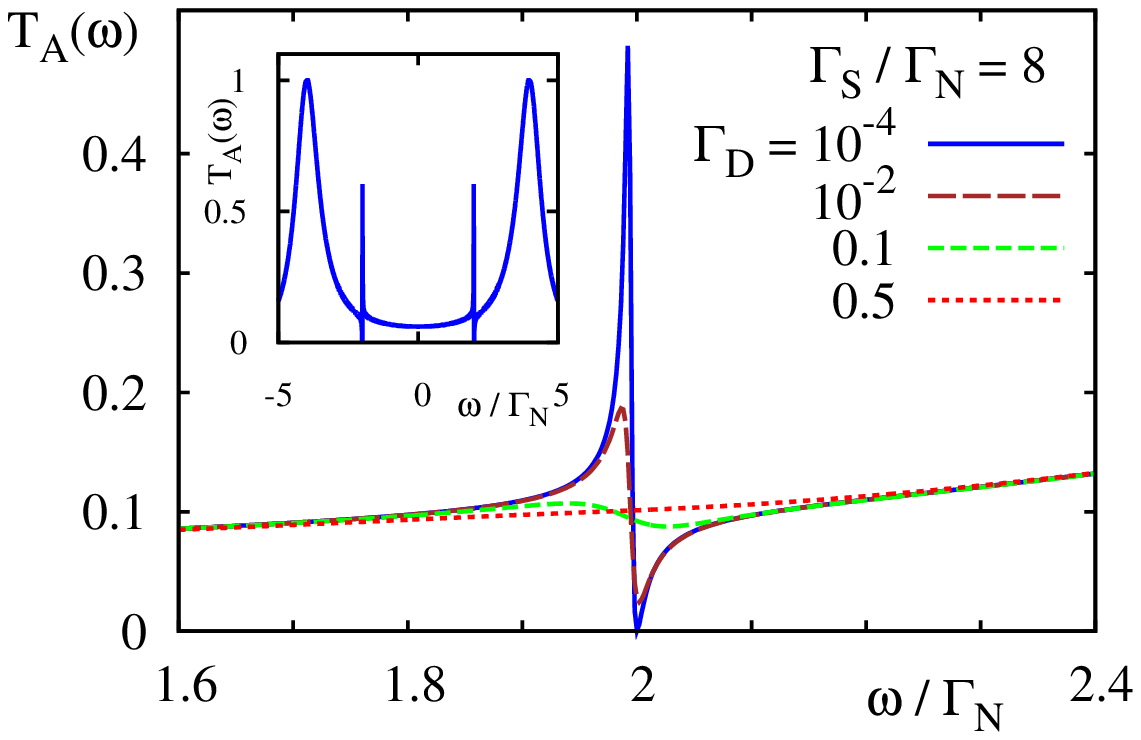}}
\caption{(color online) Andreev transmittance $T_{A}
(\omega)$ for the same parameters as in Fig.\ 
\ref{decoh_dos} ($\Gamma_{D}$ is expressed in units 
of $\Gamma_{N}$).}
\label{decoh_TA}
\end{figure}

In particular, the zero-bias differential conductance 
\begin{eqnarray} 
G_{A}(V\!=\!0) = \frac{4e^{2}}{h} \int d\omega T_{A}(\omega)
\left[ -\; \frac{d f(\omega,T)}{d \omega} \right]
\label{G_A}
\end{eqnarray} 
is at low temperatures proportional to the transmittance
\begin{eqnarray} 
G_{A}(0) = \frac{4e^{2}}{h} \; T_{A}(\omega\!=\!0) ,
\label{zeroG_A}
\end{eqnarray} 
so the optimal Andreev conductance $4e^{2}/h$ occurs when $T_{A}
(\omega)$ reaches the ideal value 1. In figure \ref{decoh_TA} 
we plot $\omega$-dependence of the Andreev transmittance for the 
same set of parameters as discussed in section III. We obtain 
the symmetric transmittance $T_{A}(-\omega)\!=\!T_{A}(\omega)$
because the anomalous Andreev scattering involves both the particle 
and hole degrees of freedom. For this reason we notice that at 
$\omega=\pm \varepsilon_{2}$ there appear the Fano-type structures 
of identical shapes but characterized by an opposite sign of 
the asymmetry parameter $q$. Again decoherence proves to have 
a detrimental influence on both these interferometric structures 
(compare the curves in Fig.\ \ref{decoh_TA} which correspond to 
several representative values of $\Gamma_{D}$).

\section{Correlation effects}

Let us now consider additional changes of the Fano lineshapes 
caused by the electron correlations. We shall restrict to the 
Coulomb repulsion at the side-attached quantum dot $U_{2}$ 
because the effects of $U_{1}$ have been already studied 
previously \cite{Baranski-11}. Briefly summarizing those studies
we can point out that the Coulomb repulsion $U_{1}$ leads to the 
charging effect and (at low temperatures) can induce the narrow 
Kondo resonance in the spectrum $\rho_{d1}(\omega)$ for $\omega 
\sim 0$. The latter effect is experimentally manifested by a slight 
enhancement of the zero-bias Andreev conductance \cite{Deacon_etal}. 
Interference effects (originating from the inter-dot coupling $t$) 
would qualitatively affect such Kondo feature if $\varepsilon_{2} 
\sim 0$. Effects of the Fano interference depend also on the 
ratio $\Gamma_{S}/\Gamma_{N}$ controlling efficiency of the 
induced on-dot pairing which competes with the Kondo physics 
\cite{Domanski-08}.

So far the correlations have been intensively studied mainly 
for the case of single quantum dot coupled between the metallic 
and superconducting electrodes \cite{Rodero-11}. For this purpose 
there have been adopted various many-body techniques, such as: 
the mean field slave boson approach \cite{Fazio-98}, noncrossing 
approximation \cite{Clerk-00}, iterated perturbative scheme 
\cite{Cuevas-01,Yamada-11}, modified slave boson method  
\cite{Krawiec-04}, numerical renormalization group calculations 
\cite{Tanaka-07,Bauer-07,Hecht-08} and other \cite{Sun-99,Cho-99,
Avishai-01,Domanski-08,Paaske-10}. The interests focused predominantly
on an interplay between the on-dot pairing and the Kondo state 
\cite{Yamada-11}. It has been experimentally  proved
\cite{Deacon_etal} that such interrelation is governed by the 
ratio $\Gamma_{S}/\Gamma_{N}$. For $\Gamma_{S} \gg \Gamma_{N}$ 
the on-dot pairing plays a dominant role (suppressing or completely 
destroying the Kondo resonance). In the opposite regime 
$\Gamma_{S} \ll \Gamma_{N}$ the Kondo state is eventually observed 
(coupling $\Gamma_{N}$ to the normal lead is necessary for that).

In this section we study the role of correlations $U_{2}$ in 
the side-coupled quantum dot taking also into account decoherence 
caused by the floating lead. For simplicity we shall neglect 
the impact of $U_{2}$ on the off-diagonal parts of ${\mb G}_{2}
(\omega)$ because the pairing induced in QD$_{2}$ for small 
interdot hopping $t$ can be anyhow expected to be marginal. 
Thus we determine the Green's function ${\mb G}_{2}(\omega)$ 
from the Dyson equation (\ref{GF}) imposing the diagonal 
selfenergy
\begin{eqnarray} 
\Sigma^{e-e}_{2}(\omega) \simeq 
\left( \begin{array}{cc}
\Sigma_{N}(\omega)  & 0 \\ 0 & 
-\left[ \Sigma_{N}(-\omega) \right]^{*} 
\end{array} \right) .
\label{approx_QD2}  
\end{eqnarray} 
Formally $\Sigma_{N}(\omega)$ denotes the selfenergy of 
the Anderson impurity immersed in the normal Fermi liquid. 
Obviously such selfenergy is not known exactly \cite{Hewson} 
therefore we have to invent some approximations.

\begin{figure}
\epsfxsize=7.5cm\centerline{\epsffile{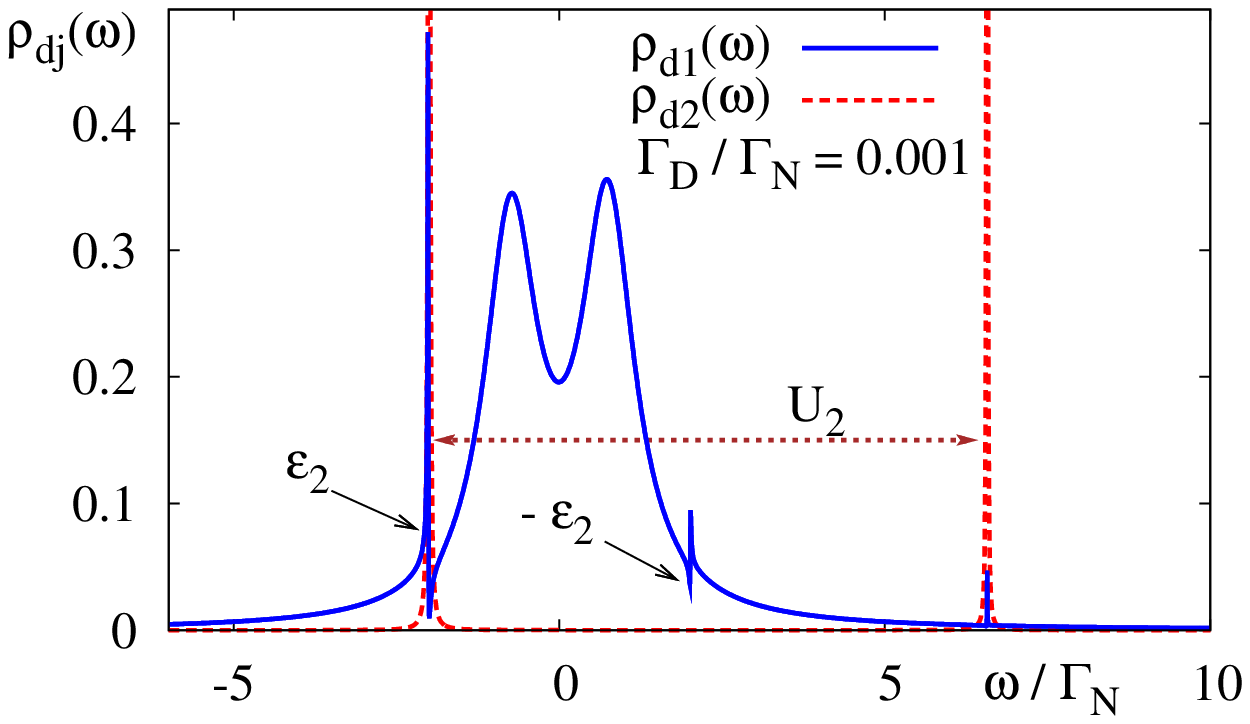}}
\vspace{-0.35cm}
\epsfxsize=7.5cm\centerline{\epsffile{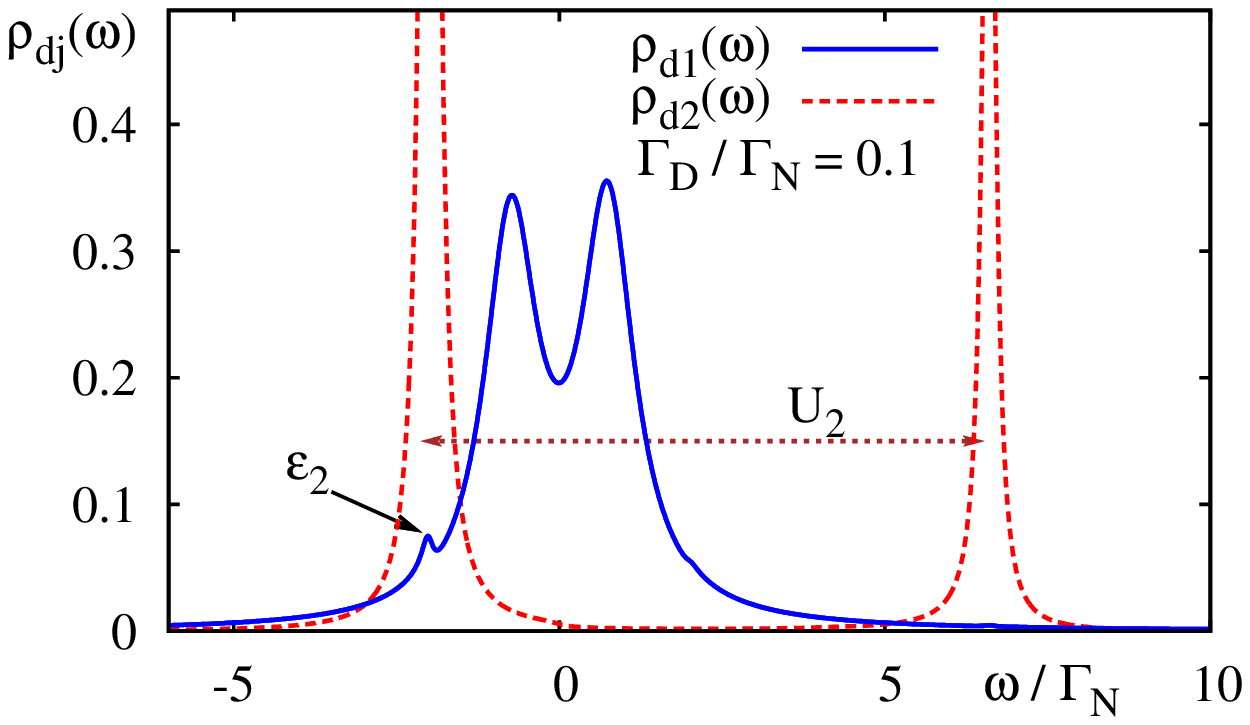}}
\vspace{-0.35cm}
\epsfxsize=7.5cm\centerline{\epsffile{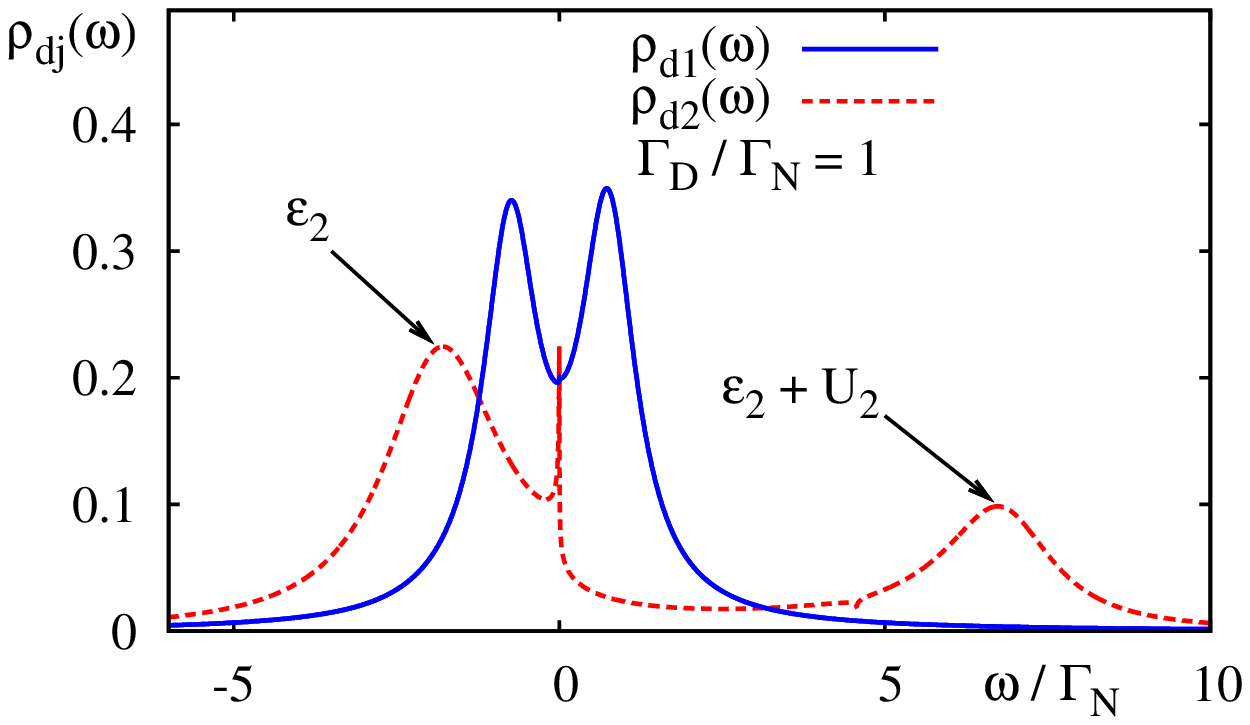}}
\caption{(color online) Evolution of the Fano-type lineshapes
for several couplings $\Gamma_{D}$ as indicated. Calculations 
have been done for $T=0.001\Gamma_{N}$ (lower than $T_{K}$), 
using the model parameters $\varepsilon_{1}=0$, $\varepsilon
_{2}=-2\Gamma_{N}$, $t=0.2\Gamma_{N}$, $\Gamma_{S}=1.5
\Gamma_{N}$ and assuming the large superconducting gap 
$\Delta=10\Gamma_{N}$.}
\label{dos_U}
\end{figure}

Among possible choices we adopt the equation of motion method 
\cite{EOM} which is capable to reproduce qualitatively the 
Coulomb blockade and the Kondo effects. Besides its simplicity 
this method is however not very precise with regard 
to the low energy structure of the Kondo peak
$\rho_{d2}(\omega \sim 0)=\frac{2}{\pi\Gamma_{D}}
\frac{T_{K}^{2}}{\omega^{2}+T_{K}^{2}}$. Nevertheless 
our results might give some hints on the qualitative trends 
and  quality of this information could be improved using
more sophisticated tools. Skipping technicalities 
discussed by us in the appendix B of Ref.\ \cite{Baranski-11} 
we can express the selfenergy $\Sigma_{N}(\omega)$ through
\begin{eqnarray}
&& \left[ \omega \!-\! \varepsilon_{2} \!-\! \Sigma_{N}
(\omega) \right]^{-1} = \label{corr_QD2} \\ && 
\frac{\tilde{\omega}-\varepsilon_{2} - [\Sigma_{N3}
(\omega)+U_{2}(1\!-\!\langle \hat{n}_{2\downarrow}\rangle)]}
{[\tilde{\omega}\!-\!\varepsilon_{2}][\tilde{\omega}\!-\!
\varepsilon_{2}\!-\!U_{2}\!-\!\Sigma_{N3}(\omega)]\!+\!U_{2} 
\Sigma_{N1}(\omega)}  \nonumber
\end{eqnarray}
where $\tilde{\omega}\!=\!\omega\!-\!\sum_{\bf k} |V_{{\bf k} D}|^{2} 
/(\omega\! - \! \xi_{{\bf k}D})\!\simeq\!\omega+\frac{i\Gamma_{D}}{2}$.
The other symbols are defined as follows $\Sigma_{N1}(\omega)\!=\!
\sum_{\bf k} |V_{{\bf k} D}|^{2} f(\xi_{{\bf k}D},T) [(\omega\! 
- \! \xi_{{\bf k}D})^{-1} + (\omega\! -\! U_{2} \! - 2 
\varepsilon_{2}\!+\!\xi_{{\bf k} D})^{-1}]$  and $\Sigma_{N3}
(\omega)\!=\!\sum_{\bf k} |V_{{\bf k} D}|^{2} [(\omega\! - \! 
\xi_{{\bf k}D})^{-1} + (\omega\! -\! U_{2} \! - 2 \varepsilon_{2}
\!+\!\xi_{{\bf k} D})^{-1}]$. This expression (\ref{corr_QD2}) for 
$\Sigma_{N}(\omega)$ substituted to the selfenergy (\ref{approx_QD2}) 
yields the Green's function ${\mb G}_{1}(\omega)$  of the central 
quantum dot via the exact relation (\ref{S_QD1}). In this way we 
can numerically determine the effect of  $U_{2}$ on $\rho_{d1}
(\omega)$ and on the Andreev transport.

For a weak interdot hopping $t$ (which is necessary to allow 
for the Fano-type quantum interference) we notice that the 
correlations $U_{2}$ can be manifested in  the spectral function
$\rho_{d1}(\omega)$ by i) the charging effect and ii) another 
characteristic structure due to the Kondo effect.

\noindent i) The first effect can be observed only if a decoherence 
is sufficiently weak, strictly speaking for $\Gamma_{D} \leq 0.1
\Gamma_{N}$. Under such circumstances the particle and hole Fano 
lineshapes (at $\pm \varepsilon_{2}$) are accompanied by two 
additional Coulomb satellites at $\pm (\varepsilon_{2}+U_{2})$. 
These interferometric features (see the top and middle panels 
of Fig.\ \ref{dos_U}) are completely washed out from the spectrum 
when $\Gamma_{D}$ slightly exceeds the value $0.1\Gamma_{N}$. This 
destructive effect of a decoherence resembles the behavior 
discussed in section III (see Fig.\  \ref{decoh_dos}) for 
the case of uncorrelated quantum dots.

\noindent ii) Instead of the particle/hole Fano lineshapes and 
their Coulomb satellites we can eventually observe a different 
qualitative structure at $\omega \sim 0$ when the coupling 
$\Gamma_{D}$ is large (provided that temperature $T < T_{K}
(\Gamma_{D})$. Its appearance is related to the Kondo resonance 
formed at the side-attached quantum dot (see the dashed curve 
in the bottom panel of Fig.\ \ref{dos_U}). Due to the inderdot 
hopping $t$ the mentioned Kondo resonance affects  the central 
quantum dot in pretty much the same way as did the 
narrow resonant level $\varepsilon_{2}$ in a weak coupling
regime $\Gamma_{D}$. Consequently we thus again observe the tiny 
Fano lineshape in the spectral function $\rho_{d1}(\omega)$ 
of the central quantum dot and in the Andreev transmittance 
$T_{A}(\omega)$ near $\omega \sim 0$. 

\begin{figure}
\epsfxsize=8.5cm\epsffile{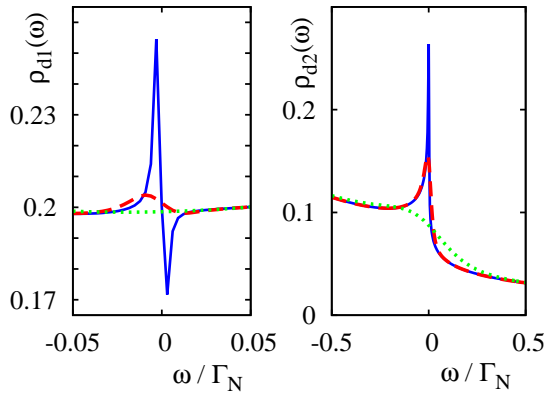}
\caption{(color online) Influence of the Kondo effect appearing
in  $\rho_{d2}(\omega)$ on a tiny  Fano-type structure of 
the central quantum dot spectrum $\rho_{d1}(\omega)$ near 
$\omega=0$.  The curves have been calculated using the same 
parameters as in figure \ref{dos_U} for the following temperatures
$T/\Gamma_{N}=0.001$ (solid line), $0.01$ (dashed line) and 
$0.1$ (dotted line).}
\label{Kondo_effect}
\end{figure}

Since the Kondo-induced interferometric structure is hardly 
noticeable on the large energy scale we show it separately 
in Fig.\ \ref{Kondo_effect} restricting to a narrow regime 
around the Fermi level $\omega=0$. Let us remark that the
Kondo resonance in $\rho_{d2}(\omega)$ and its Fano-type 
manifestation in $\rho_{d1}(\omega)$ are both very sensitive 
to temperature. This fact proves that the considered Fano 
lineshape at $\omega \sim 0$ is intimately related to the Kondo 
effect on the side-attached quantum dot.

\section{Conclusions}

In summary, we have investigated the influence of decoherence
and electron correlations on the interferometric Fano-type 
lineshapes of the double quantum dot coupled in T-shape configuration 
to the conducting and superconducting leads. We find evidence that
already a weak decoherence can consequently smear out the Fano 
lineshapes of the particle and hole states. On a microscopic 
level this detrimental influence can be assigned to a broadening 
of the resonant levels near $\pm \varepsilon_{2}$, so that the 
phase shift of the scattered electron waves is no longer sharp 
and therefore the Fano-type interference cannot be satisfied 
\cite{Miroshnichenko-10}.

The correlations $U_{2}$ on the side-attached quantum dot 
have the additional qualitative influence. For a weak decoherence
the particle/hole Fano structures at $\pm \varepsilon_{2}$ are
accompanied by appearance of their Coulomb satellites at $\pm (\varepsilon_{2}
+U_{2})$. All these interferometric features gradually disappear
upon increasing $\Gamma_{D}$ (i.e.\ for stronger decoherence).
On the other hand, in the opposite regime of strong coupling 
$\Gamma_{D}$, the narrow Kondo resonance appears in the spectral 
function $\rho_{d2}(\omega)$ of the side-coupled quantum dot. 
Its formation gives rise to the new interferometric structure 
appearing in the spectrum of the central quantum dot at 
$\omega \sim 0$. This temperature dependent Fano-type 
lineshape is observable in the spectral function $\rho_{d1}
(\omega)$ and would be detectable in the Andreev conductance. 
Such Kondo-induced Fano effect is however very tiny therefore 
its experimental verification might be challenging. 

\begin{acknowledgments}
We acknowledge useful discussions with B.R.\ Bu\l ka and 
K.I.\ Wysoki\'nski. This project is supported by the National 
Center of Science under the grant NN202 263138.
\end{acknowledgments}

\appendix
\section{Gap edge features}

\begin{figure}
\epsfxsize=10cm\centerline{\epsffile{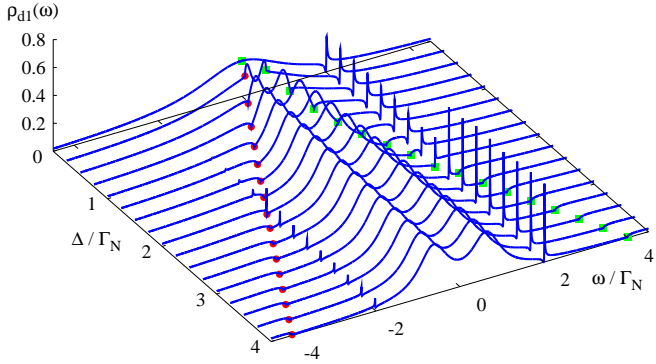}}
\caption{(color online) Evolution of the particle/hole Fano 
lineshapes at $\pm \varepsilon_{2}$ obtained for $\varepsilon_{1}=0$, 
$\varepsilon_{2}=2\Gamma_{N}$, $t=0.2\Gamma_{N}$, $\Gamma_{S}
=1.5\Gamma_{N}$, $U_{i}=0$ and for the varying magnitude of
$\Delta$. The filled circles and squares indicate the cusp-like 
signatures of the gap edge singularites at $-\Delta$  and 
$+\Delta$.}
\label{dos_Delta}
\end{figure}

There is also another important energy scale, relevant for 
the present study. It is related to a magnitude of the energy 
gap $\Delta$ of superconducting lead. To illustrate its influence 
on the spectral function $\rho_{d1}(\omega)$ we show in figure 
\ref{dos_Delta} variation within the region $0 \leq \Delta \leq 4\Gamma_{N}$.
If the energy gap is small we see that the proximity effect is very
fragile. For this reason we hardly notice the Fano-type structure at 
$-\varepsilon_{2}$ because the on-dot pairing is rather ineffective. 
The Fano resonance starts to be well pronounced  at $-\varepsilon_{2}$
when $\Delta$ becomes comparable (or larger) than $\Gamma_{S}$. 
Additionally, the energy gap $\Delta$  is responsible for two tiny 
dips appearing at $\omega = \pm \Delta$. They are signatures of 
the gap edge singularities of superconducting lead. 
Roughly speaking, outside the energy regime $|\omega| > \mbox{\rm min}
\left\{ \Delta, \Gamma_{S}/2 \right\}$ the charge tunneling occurs 
via the usual single particle channel and the Andreev tunneling is 
there no longer dominant \cite{Fazio-98,Clerk-00,Cuevas-01,
Krawiec-04,Sun-99,Domanski-08}.

\end{document}